\documentstyle[sprocl,latexsym,amssymb,epsfig,wrapfig,bm]{article}

\def\NPB{{\em Nucl. Phys.} B}

\def\EPJ{{\em Eur. Phys. J.} C }
 \newcommand{\fr}[2]{\frac{{\displaystyle #1}}{{\displaystyle #2}}}

    {\end{list}}

\def\be{\begin{equation}}
\def\ee{\end{equation}}
\def\bea{\begin{eqnarray}}
\def\eea{\end{eqnarray}}
\newcommand{\ggam}{\mbox{$\gamma\gamma\,$}}
\newcommand{\ggww}{\mbox{$\gamma\gamma\to W^+W^-\,$}}

\begin{document}

\title{CHARGE ASYMMETRY IN $\bm{\gamma\gamma \to \mu^+\mu^- + neutrinos}$
WITH POLARIZED PHOTONS}

\author{ D.~A.~ANIPKO$^a$, M.~CANNONI$^b$, I.~F.~GINZBURG$^a$,
A.~V.~PAK$^a$, O.~PANELLA$^b$}

\address{
$^a$ Sobolev Institute of Mathematics, Novosibirsk, 630090, Russia,\\
$^b$ LPNHE, 4 Place Jussieu, 75525 Paris, France\\
$^c$ INFN, Sezione di Perugia, Via G. Pascoli 1, 06129, Perugia, Italy }

\maketitle

\abstracts{ The difference in distributions of $\mu^+$ and $\mu^-$
in reactions $\gamma \gamma \to \mu^+\mu^-+\nu\bar\nu$ and $\gamma
\gamma \to W^\pm\mu^\mp +\nu(\bar\nu)$ with polarized photons at
$\sqrt{s}>200$ GeV is a large observable effect which is sensitive
to New Physics phenomena.}

The Photon Collider~\cite{Ginzburg1} option of the next generation
linear colliders (LC)~\cite{Tesla} offers the opportunity to study
with high precision the physics of gauge bosons. The photons with
largest energy will be produced mainly in states with
definite helicity $\lambda_i\approx \pm1$.
The Standard Model (SM) cross section of $\gamma \gamma \to W^+W^-$ process at
energies greater than $200$ GeV is about $80$ pb \cite{GKPS} and
ensures very high event rates. We study the charge asymmetry (CA)
of leptons produced in this process in the SM and make preliminary
considerations on how these asymmetries change due to some
possible effects of New Physics. In this note we present results
of Ref.~\cite{1stpub} and some results obtained after that publication.

\begin{wrapfigure}[15]{l}[0pt]{5.5cm}
\includegraphics[height=4cm,width=5cm]{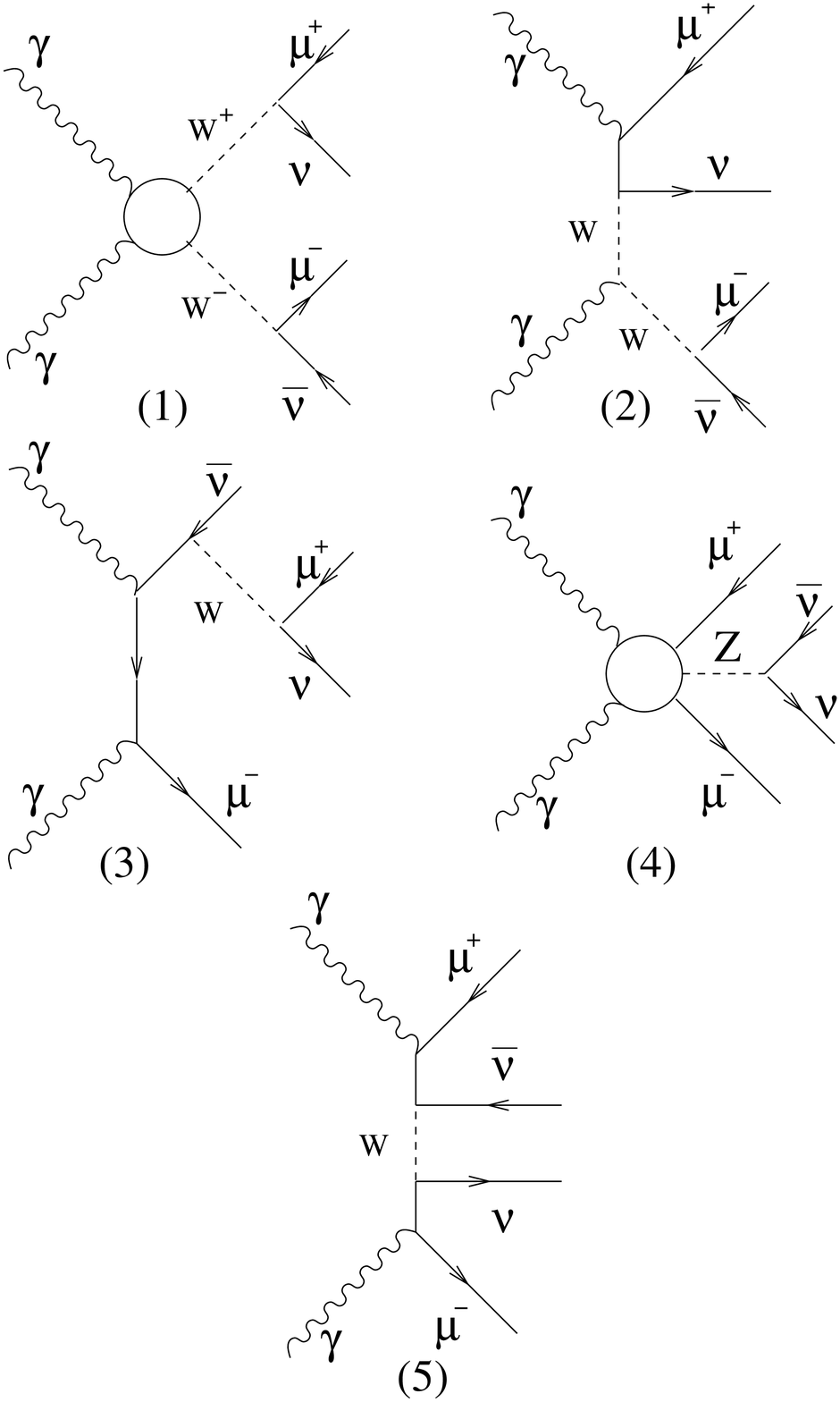}
\caption{Classes of tree level contributing Feynman diagrams.}
\label{diag}
 \end{wrapfigure}

$\blacksquare$ {\bf Diagrams. Qualitative description.} In SM, at
the tree level, the process $\gamma \gamma \to \mu^+\mu^-
\nu\bar\nu$ is described by 19 diagrams, subdivided into five
classes, shown in Fig.~\ref{diag}. The collection of diagrams
within each class is obtained from those shown in the figure with
the exchange $+ \leftrightarrow -$, $\nu\leftrightarrow \bar{\nu}$
and permutations of
photons. For each class we first give an estimate of its
contribution to the total cross section, based on the equations
for $2\to 2$ processes at $s\gg M_W^2$, assuming for the SM gauge
couplings $g^2\sim g'^2\sim\alpha$ and denoting $B=Br(W\to\mu\nu)$
and $B_Z=Br(Z\to \nu\bar{\nu})$:

(1) 3 double-resonant diagrams (DRD) describe $WW$ production and
decay, $ \sigma_1 \sim (\alpha^2/M_W^2) B^2\approx \sigma_{tot}$;

(2) 4 single-resonant diagrams  with $W$ exchange in $t$--channel,
$\sigma_2\sim (\alpha^3/M_W^2)B\sim \alpha\sigma_d/B\sim
0.05\;\sigma_{tot}$. This contribution is eliminated almost
completely by its interference with DRD;

(3) 4 single resonant diagrams with $\mu$ exchange in $t$--channel
(gauge boson bremsstrahlung), $\sigma_3\sim (\alpha^3/s)B\sim
\alpha\sigma_dM_W^2/(Bs)<0.01\sigma_{tot}$;

(4) 6 diagrams with radiation of $Z$ boson in the process
$\gamma\gamma\to\mu^+\mu^-$, $\sigma_4 \sim (\alpha^3/s)B_Z\sim
\alpha\sigma_dM_W^2B_Z/(B^2s)<0.01\sigma_{tot}$;

(5) 2 non-resonant diagrams, $\sigma_5\sim \alpha^4/M_W^2\sim
\alpha^2\sigma_dM_W^2/(B^2s)\ll 0.01\sigma_{tot}$.

The CA is present in the DRD contribution, and simulation shows
that the contributions of other diagrams are negligible.

{\it We now give a qualitative discussion of the origin of CA
referring to the dominant DRD contribution, \ggww.} ({\it i}) The
cross section practically does not depend on photon polarizations.
({\it ii}) The $W$ bosons are distributed around the forward and
backward directions~\cite{GKPS}, $d\sigma\propto
1/(p_\bot^2+M_W^2)^2$. ({\it iii}) The helicity of $W$ moving in
the positive direction is $\lambda_{W_1}\approx \lambda_1$, while
$\lambda_{W_2}\approx \lambda_2$,  independently on the charge
sign of $W$ (helicity conservation,~\cite{BBB}). ({\it iv}) Let
the $z'$--axis be directed along the $W$ momentum $\vec{p}_W$,
$\varepsilon\approx M_W/2$ and $p_{z'}$ the energy and the
longitudinal momentum of $\mu$ in the $W$ rest frame. The
distribution of muons from the decay of $W$  with charge $e=\pm 1$
and helicity $\lambda=\pm 1$ in its rest frame is
$\propto(\varepsilon -e\lambda p_{z'})^2$. Hence, the distribution
of muons from the decay of $W^\pm$ has a peak along $\vec{p}_W$ if
$e\lambda_W=-1$ and opposite to $\vec{p}_W$ when $e\lambda_W=+1$.
These distributions are boosted to the $\gamma\gamma$ collision
frame. For example, if both colliding photons have $\lambda = -1$,
the produced $\mu^+$ are distributed around the upper value of
their longitudinal momentum (both in the forward and backward
direction), while $\mu^-$ are concentrated near the zero value of
their longitudinal momentum. This boost makes the distribution in
$p_\bot$ wider in the first case and narrower in the second case.

$\blacksquare$ {\bf Numerical results} have been obtained with the
CompHEP package~\cite{Pukhov} considering the complete set of
diagrams. The following cuts for background suppression are
applied: a cut on the muons scattering angles given by
$\pi-\theta_0> \theta > \theta_0$, with $\theta_0=10$ mrad; a cut
on muons transverse momentum $p_\perp>10$ GeV, both on each muon
and on the couple of muons.
\begin{figure}[t!]
\begin{center}
\includegraphics[clip,height=3cm,width=4.6cm]{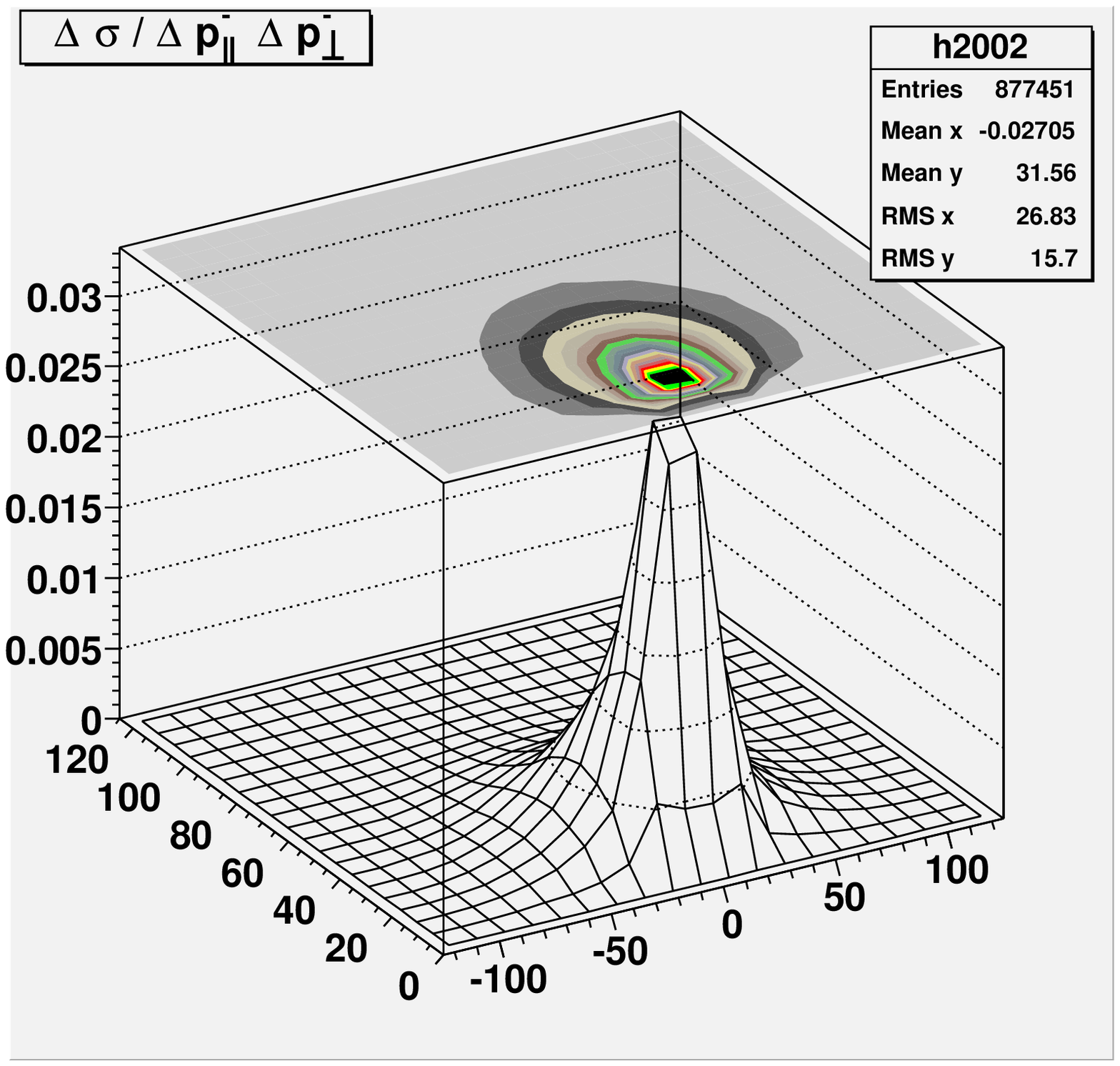}\hspace{1cm}
\includegraphics[bb=0 30 473 467,clip,height=3cm,width=4.6cm]{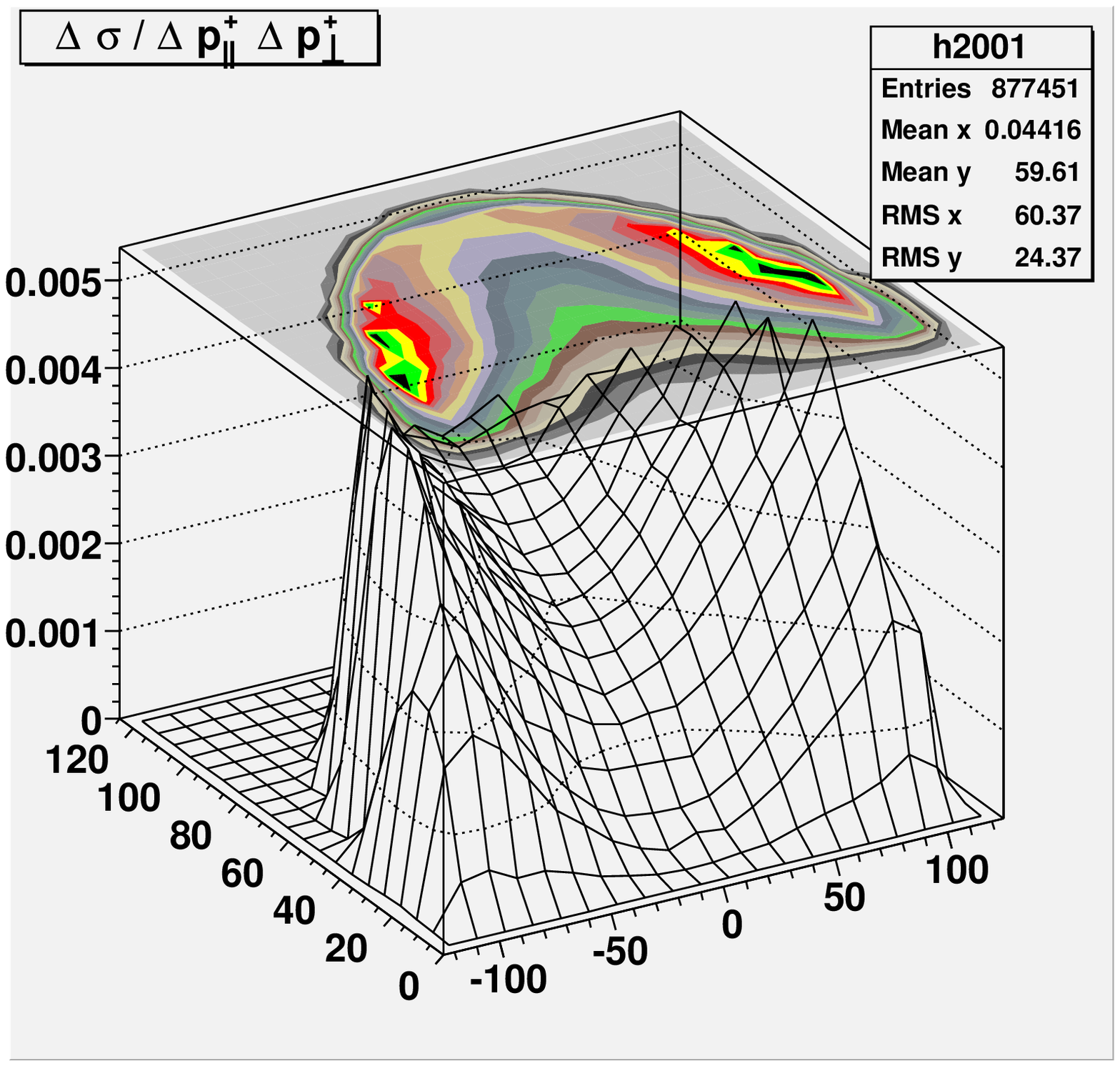}\\
\includegraphics[clip,height=3cm,width=4.6cm]{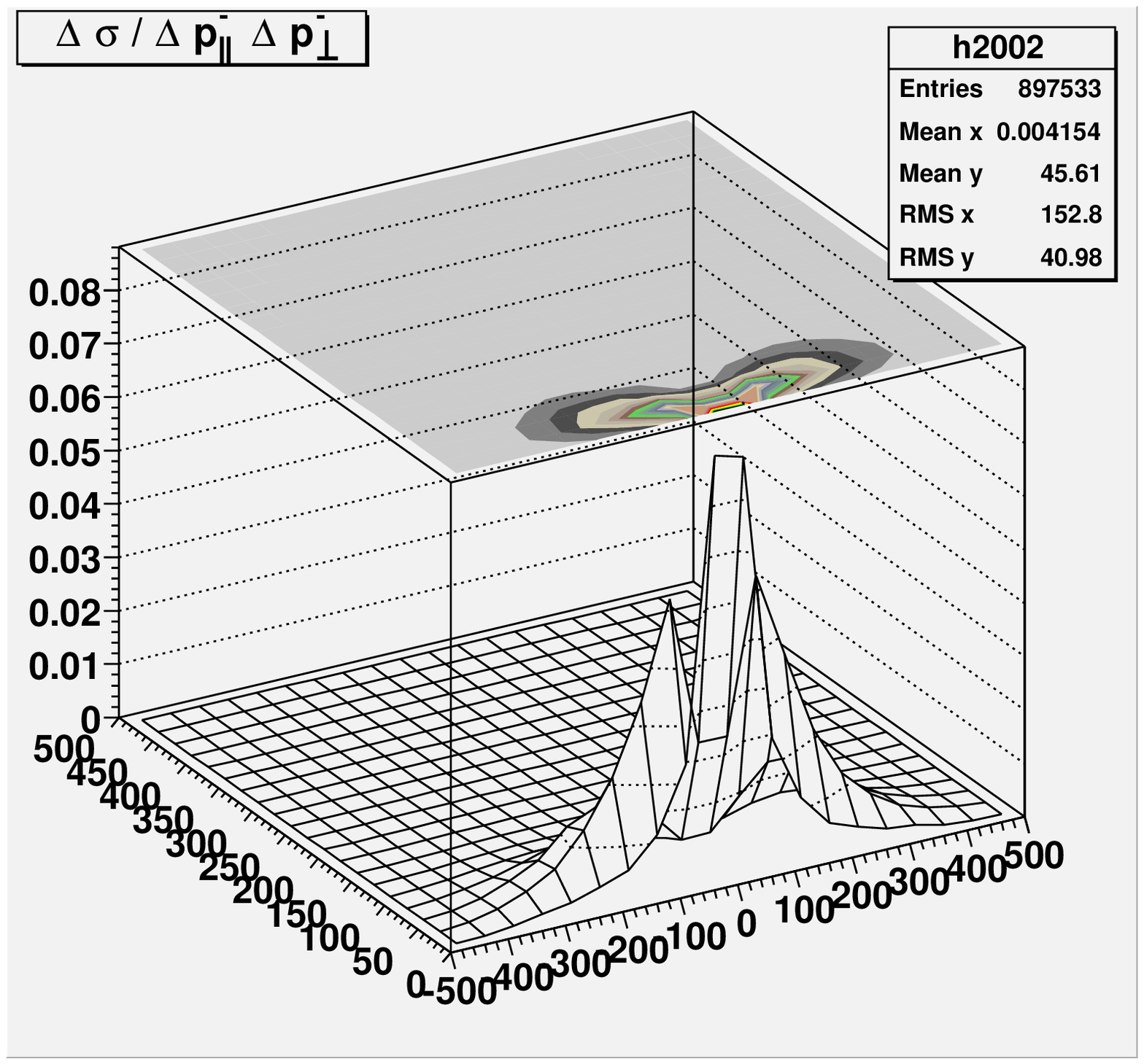}\hspace{1cm}
\includegraphics[clip,height=3cm,width=4.6cm]{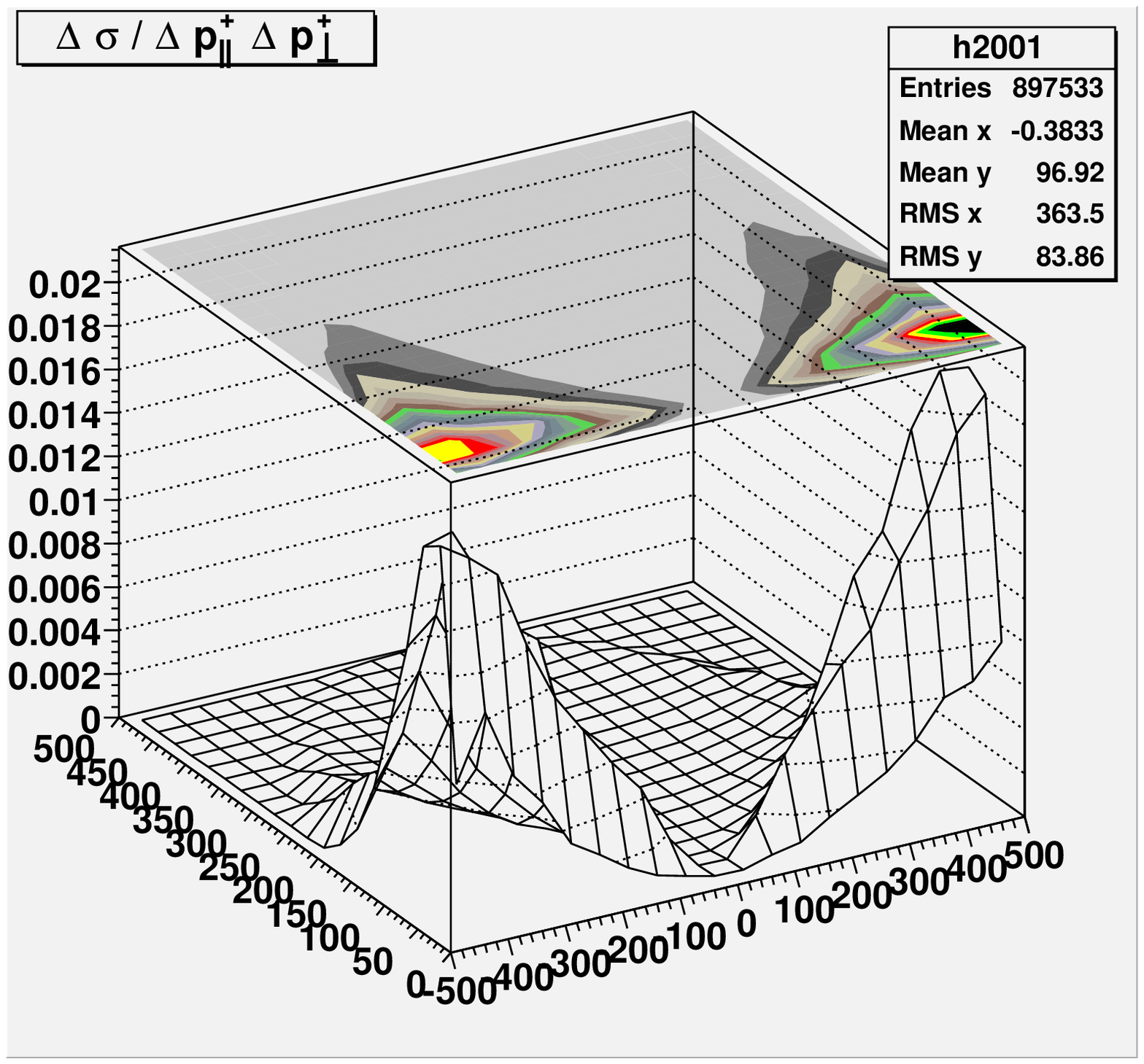}
\caption{ Distributions for $(-\, -)$ helicity of colliding
photons: $\mu^-$ on the left, $\mu^+$ on the right:
$\sqrt{s_{\ggam}}=250$ GeV  (top) and $\sqrt{s_{\ggam}}=1000$ GeV
(bottom).}
\label{energyvar}
\end{center}
\end{figure}
Fig.~\ref{energyvar} shows the distributions of muons over longitudinal $p_\|$ and
transverse $p_\perp$ components of muon momentum $\partial^2
\sigma/(\partial p_{\parallel}\partial p_{\perp})$ at different
energies for the monochromatic beams. It shows clearly {\em a
strong difference in the distributions of $\mu^-$ and $\mu^+$}.
The absolute value of the effect decreases with energy due to the
increasing importance of the applied cuts.
\begin{figure}[t!]
\begin{center}
\includegraphics[height=3cm,width=4.6cm]{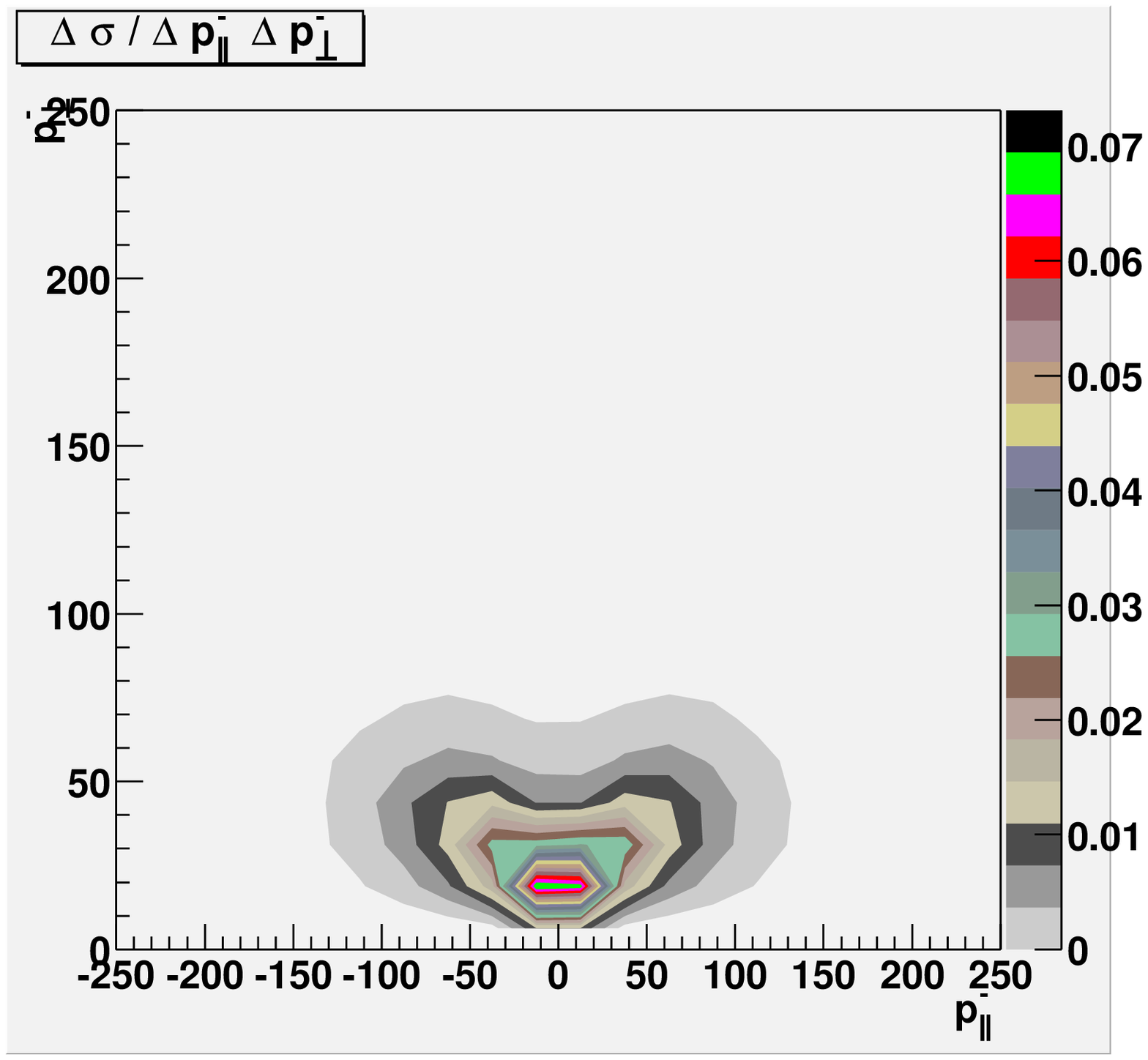}\hspace{1cm}
\includegraphics[height=3cm,width=4.6cm]{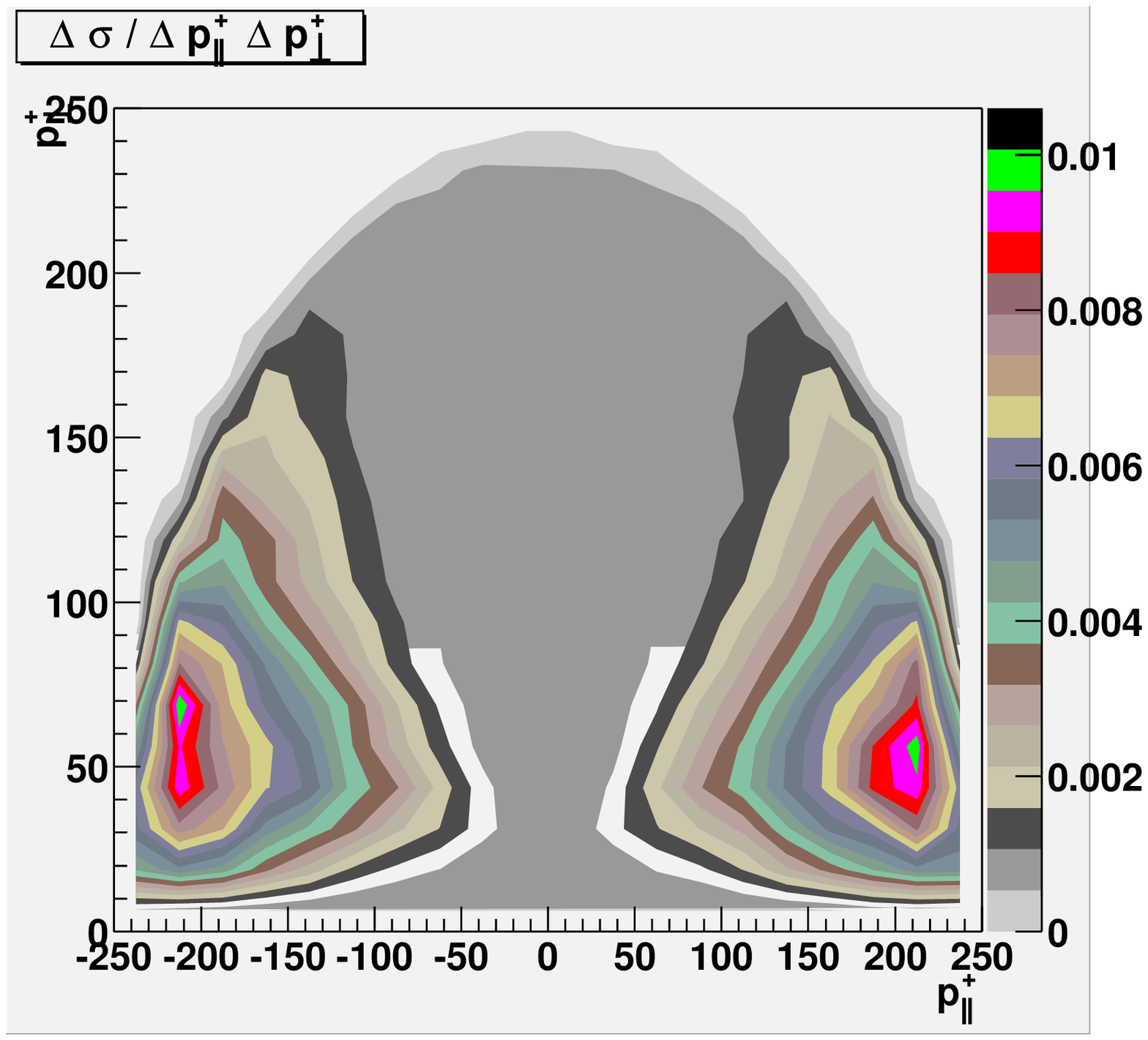}\\
\includegraphics[height=3cm,width=4.6cm]{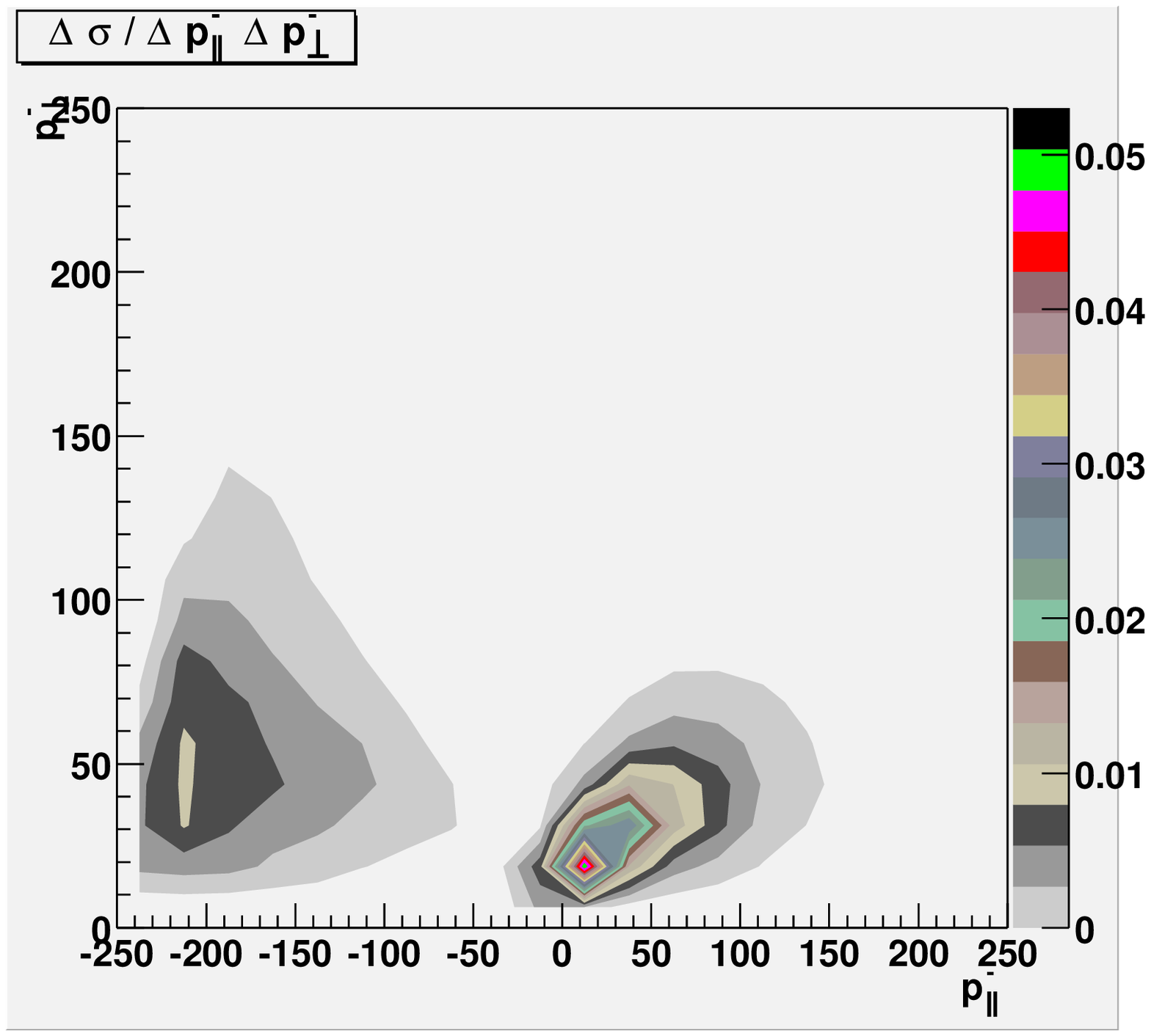}\hspace{1cm}
\includegraphics[height=3cm,width=4.6cm]{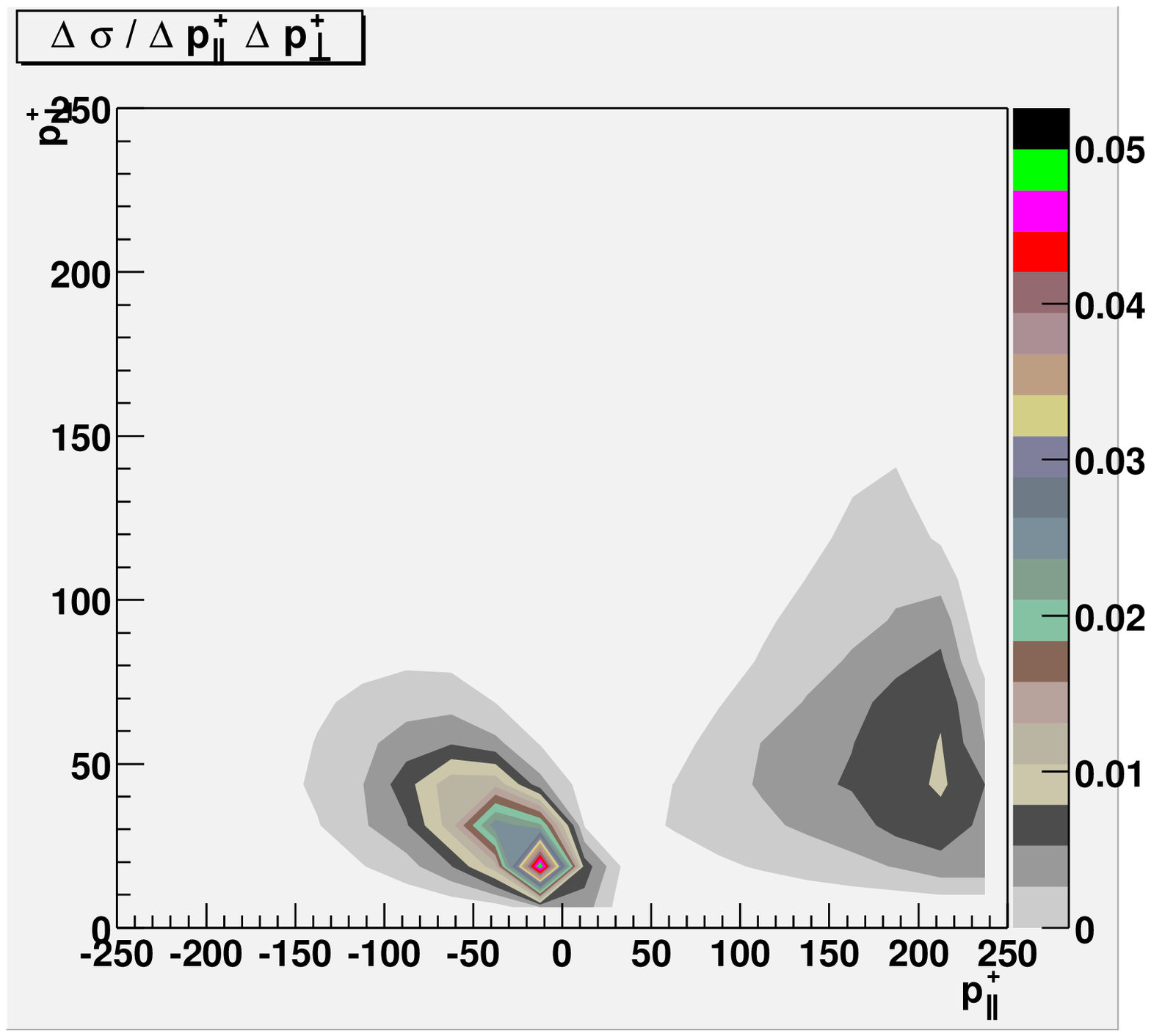}
\caption{ Distributions in the
$(p_\parallel,p_\perp)$ plane (pb/bin). Left -- $\mu^-$, right --
$\mu^+$; helicities of colliding photons:  $(-\,-)$ -- top,
$(-\,+)$ -- bottom, $\sqrt{s_{\ggam}}=500$ GeV. }
\label{mm}
\end{center}
\end{figure}
In Fig.~\ref{mm} we show the same distributions in the form of
two-dimensional level lines, for different initial helicity
states. Note that the scale of effect in the graphs with $\mu^-$
for $(-\,-)$ case differ from other cases by a factor of seven.
Due to CP conservation, the $\mu^\pm$ distributions for $(-\,-)$
case coincides with $\mu^\mp$ distribution for $(+\,+)$ case. The
obtained distributions in the $(p_\|,p_\bot)$ plane have the form
which corresponds to the qualitative picture outlined above.

In Ref.~\cite{1stpub} we also characterized CA by the quantities
$\Delta_{L,T}= \fr{P_{L,T+}^- -P_{L,T+}^+}{P_{L,T+}^-
+P_{L,T+}^+}$ calculated by the normalized mean values of
longitudinal and transverse momentum of muons flying in the
forward hemisphere, $P_{L,T+}^\pm=\fr{\int p_{\|,\perp}^\pm
d\sigma}{ E_{\gamma max}\int d\sigma}$ and shown that these
quantities change only weakly both using "realistic" photon
spectra~\cite{GinK} and with the addition of the anomalous
magnetic moment of $W$. The quadruple moment of $W$ changes
$\Delta_L$ significantly at least in the $(-\,-)$ case: this
asymmetry can be useful for the study of the $\lambda$ anomaly. To
mimic the effect of new interactions and/or new particles, we
considered a toy model with a "muon" having a mass of 40 GeV.
Essential variations are found in this case so that the study of
CA can be a useful tool for the discovery of new particles.

$\blacksquare$ {\bf Outlook}. The effects considered so far are
identical for electrons and muons, thus the same asymmetry will be
observed in all $\ell^+\ell^-$ distributions with $\ell=e$ or
$\mu$. Therefore, all these contributions should be gathered for a
complete analysis. This will enhance the value of the observable
cross section for $\gamma\gamma\to\ell^+\ell^-\nu\bar{\nu}$  from
0.9 pb for to 3.7 pb and for $\gamma\gamma\to W^+\ell^-\bar{\nu}$,
etc. to 23.5 pb (millions of events per year). The observable
final state receives contributions also by processes like
$\gamma\gamma\to\mu^+\mu^-\nu\bar{\nu} + \nu\bar{\nu}$ pairs. For
example, the most important one will be
$\gamma\gamma\to\tau^+\mu^- \nu\bar{\nu}$, etc., with subsequent
decay $\tau\to \mu \nu\bar{\nu}$. The cross section of this
process is 17\% of those discussed above ($Br(\tau\to \mu
\nu\bar{\nu})$ + 17\% for the case with the change $\tau^+\to
\tau^-$, etc. +3\% for $\gamma\gamma\to\tau^+\tau^-\nu\bar{\nu}$).
We plan to consider these qualitatively discussed channels in near
future.\\

This work was supported by grants RFBR 02-02-17884,
NSh-2339.2003.2,  grant 015.02.01.16 Russian Universities and by
the European Union under contract N. HPRM-CT-2002-00311.
M.~C. thanks ``Fondazione Angelo Della Riccia'' for a fellowship 
and C.~Carimalo and LPNHE for the kind hospitality.

\end{document}